\documentclass[ twocolumn, showpacs, pra, superscriptaddress, floatfix
 amsmath,amssymb,
 aps,
floatfix,
]{revtex4-1}
\usepackage{CJK}

\usepackage{graphicx}
\usepackage{dcolumn}
\usepackage{comment}
\usepackage{xcolor}
\usepackage[colorinlistoftodos]{todonotes}
\usepackage{hyperref}
\usepackage[english]{babel}
\usepackage{bm}
\usepackage{amsmath}
\usepackage{amssymb}
\usepackage{comment}
\usepackage{subfigure}
\usepackage{epstopdf}
\usepackage{hyperref}
\usepackage{appendix}
\usepackage{float}
\usepackage{booktabs}

\usepackage{mathtools}

\DeclarePairedDelimiterX\braket[2]{\langle}{\rangle}{#1 \delimsize\vert #2}

\UseRawInputEncoding

\makeatother

\newcommand{\Rnum}[1]{\uppercase\expandafter{\romannumeral #1\relax}}

\begin{document}


\title{Plasma effects on lifetimes and screening of Rydberg excitons}

\author{AbdAlGhaffar K. Amer}
 \email{amer1@purdue.edu}
 \affiliation{
 Department of Physics and Astronomy, Purdue University, West Lafayette, Indiana 47906 USA
}

\author{V. Walther}%
 \email{vwalther@purdue.edu}
\affiliation{
 Department of Chemistry, Purdue University, West Lafayette, Indiana 47906 USA
}
 \affiliation{
 Department of Physics and Astronomy, Purdue University, West Lafayette, Indiana 47906 USA
}

\author{F. Robicheaux}%
 \email{robichf@purdue.edu}
\affiliation{
 Department of Physics and Astronomy, Purdue University, West Lafayette, Indiana 47906 USA
}

\begin{abstract}
We simulate the effects of a neutral electron--hole plasma on Rydberg excitons in cuprous oxide (Cu$_2$O), focusing on the validity of Debye screening and the role of plasma-induced thermalization. Unlike atomic Rydberg states, excitons in Cu$_2$O consist of quasiparticles with comparable effective masses whose orbital frequencies can exceed the plasma frequency, invalidating the assumption of a stationary screened charge. Using two complementary approaches, a classical orbit model and a harmonic-oscillator representation evolved via the truncated Wigner approximation, we study exciton lifetimes and interaction screening under realistic plasma conditions. We find numerically that plasma-induced scattering 
induces finite exciton lifetimes with specific scaling relations with plasma density, principal quantum number $n$ and temperature,
possibly providing an explanation for experimentally observed deviations from the $n^3$ scaling at high principal quantum numbers. By explicitly computing time-averaged electric fields, we show that Debye screening overestimates the screening of the exciton's internal field, especially for high angular momentum states. 
Furthermore, we demonstrate that exciton-exciton interactions are not Debye screened at separations comparable to the Debye length for Rydberg excitons that are well resolvable in absorption measurements. 
\end{abstract}

\maketitle

\section{\label{sec:level1}Introduction}

Excitons in semiconductors are bound electron-hole pairs that serve as a condensed matter analogue of hydrogen-like atoms. In cuprous oxide (Cu$_2$O), Rydberg excitons with high principal quantum number $n \sim 30$ were observed \cite{kazimierczuk2014giant, versteegh2021giant}, with electron-hole separations up to a micrometer. These Rydberg excitons exhibit features similar to atomic Rydberg states, such as the dipole-dipole blockade effects \cite{kazimierczuk2014giant,bergen2023large,zielinska2025quantum,heckotter2020experimental}. Due to their large sizes and polarizabilities, Rydberg excitons are highly susceptible to charges in the crystal environment. Bound charges emerge as vacancy centers, exerting a distribution of electric fields in the semiconductor \cite{Krueger2020a}, whereas free charge carriers can be introduced through pumping or gating, but are even thought to be present in neutral ground semiconductors in their ground state. 

There is a long history of combined theoretical and experimental work to describe the effects of plasma on excitonic ground and Rydberg states \cite{glazov2018, klingshirn2007semiconductor}. One of the most basic descriptions is the classical Debye-H\"{u}ckel model (shortened as Debye model) of plasma screening. 
This model assumes a stationary positive charge (analogous to an immobile nucleus), around which the mobile charges arrange according to the thermal distribution. However, unlike atomic Rydberg states that have massive nuclei, both electron and hole forming the Rydberg excitons have comparable effective masses and consequently exhibit comparable motion. This motion can occur on timescales faster than the plasma's oscillation periods depending on the energy level and the plasma density. Although the Debye model successfully captures the main observed features of Rydberg excitons in Cu$_2$O \cite{heckotter2018rydberg, walther2020plasma}, the lack of a separation of time scales raises questions about the direct applicability of the Debye model for Rydberg excitons. 

In addition, theoretical and experimental studies of the Rydberg series in Cu$_2$O \cite{stolz2022scrutinizing,seidel1995energy,arndt1996two,semkat2019influence,semkat2021quantum,stolz2021coherent} showed that the Debye model can yield 
inconsistent results when used to simultaneously explain the band gap shift and the energy shifts of Rydberg excitons. Specifically, the ratios of plasma density to temperature required for the Debye model to reproduce the observed band edge shift are approximately an order of magnitude higher than those required to fit the exciton energy shifts. Although the quantum many-body approach developed in these studies provides a closer fit to the experimental data, it does not explain the physical origin of the Debye model's failure in this regime \cite{stolz2022scrutinizing}. In parallel with these questions of screening, recent experiments have revealed deviations from the expected $n^3$ scaling of Rydberg exciton lifetimes at high principal quantum numbers \cite{kazimierczuk2014giant,heckotter2020experimental}. While phonon-induced decay channels account for the dominant lifetime contributions at low $n$, the origin of the reduced lifetimes at higher $n$ remains an open question. 

Another question concerns the screening of \emph{inter-}particle interactions in the presence of a plasma. A theoretical analysis of atomic neutral plasmas showed that free charges generally produce only partial (non-exponential) screening of long-range inter-excitonic interactions, i.e., that van der Waals forces retain their characteristic $r^{-6}$ tail at large interparticle distances $r$ \cite{alastuey2007van}. This predicts a non-vanishing  dipolar Rydberg blockade should persist in neutral plasmas.

Motivated by these observations, the present work aims to address three closely related questions. First, we investigate whether plasma-induced scattering can account for the observed deviations from the expected lifetime scaling of Rydberg excitons in Cu$_2$O. Second, we assess the validity of Debye screening for excitons whose internal dynamics are fast compared to the plasma response. Third, we examine the plasma screening of the exciton-exciton interactions. We address these questions using classical and semiclassical models for both the plasma and the exciton to provide clear physical insights into the underlying mechanism.

In Sec. \ref{sec:level2}, we outline our theoretical framework. In Sec. \ref{methods_plasma_num}, we discuss our model of the plasma and the numerical techniques used for its simulation. In Sec. \ref{subsec_ExcitonModels}, we describe two complementary exciton models, and in Sec. \ref{methods_wigner_num}, we detail the method used to evolve the exciton state. The results section, Sec. \ref{sec_Results} is split into four parts. In Sec.\ref{Sec_thermalneo}, we discuss the effects of the plasma on the lifetime of the exciton's states and how they depend on the plasma density and temperature. In Sec. \ref{reslts_screening}, we discuss the limitations of the Debye screening model on a classical orbit of a bound electron-hole pair. Section \ref{Sec_thermScreen} merges the analysis for the screening and the lifetimes in order to quantify the extent of observable plasma screening for the exciton states. Finally in Sec. \ref{results_coupled}, we analyze the plasma effects on exciton-exciton interactions.

\section{\label{sec:level2}Methods}

We consider a many-body system comprising a neutral plasma of free electrons and holes, together with one or several excitons in Rydberg states.
A comprehensive quantum-mechanical description of this system is provided by the dynamically screened Bethe-Salpeter equation for the electron-hole pair \cite{semkat2019influence}. In practice, an exact solution is generally intractable due to many-body effects induced by the plasma, including Pauli blocking, exchange interactions, and the dynamical screening of the electron-hole interaction. Established approximate approaches are reviewed in Refs.~\cite{haug1984electron, glazov2018}.
Here, we focus on experimentally relevant regimes for excitons in Cu$_2$O, where plasma densities are low \cite{heckotter2018rydberg}. In this limit, the interparticle spacing of the free charges is sufficiently large that quantum correlations between electrons and holes can be neglected.
Rather than approximating the dynamically screened Bethe-Salpeter equation directly, we adopt an ad hoc semiclassical model in which the unbound electrons and holes are treated as classical point charges moving under the total electric forces in the system.

An advantage of this approach is that it retains aspects of dynamical screening, leading to time-dependent perturbations and, consequently, a finite lifetime of the bound electron–hole states; effects that are absent in quasi-static treatments. We expect this classical plasma description to be reliable as long as the plasma is weakly correlated.
In principle, one could then solve the Schr\"odinger equation for the electron-hole pair in the presence of this uncorrelated, time-dependent plasma environment, potentially including small deviations from a quadratic band structure \cite{Krueger2020a}.

Instead, to further simplify the description of the Rydberg exciton, we employ two complementary models. In the first, fully classical approach, the exciton is treated as a Keplerian two-body system evolving under classical forces. In the second, semiclassical approach, the internal degrees of freedom of the exciton are mapped onto a quantum harmonic oscillator.

In this section, we outline the theoretical framework used to model these components. Section \ref{methods_plasma_num} details the numerical simulation of the plasma background and shows that it reproduces the known Debye screening for stationary point charges. Section \ref{subsec_ExcitonModels} details the two models used to represent the exciton and its coupling to the plasma. Section \ref{methods_wigner_num} describes the numerical method to time evolve the system via the Truncated Wigner Approximation (TWA). 


\subsection{\label{methods_plasma_num} Numerical simulation of the plasma} 
In this section, we briefly discuss the numerical technique we utilize to simulate the plasma part of the system and show that it reproduces the known Debye screening for point charges. For initialization, the positions of the plasma particles are randomly placed with equal probability inside a sphere with radius $R$ such that the two-species density of the plasma particles is $\rho$, with equal electron and hole densities $\rho_e = \rho_h = \rho/2$. Thus the plasma frequency \cite{chen1984introduction} becomes
\begin{equation}
\label{eq_plasma_freq}
    \omega_p = \sqrt{\frac{\rho e^2}{2 \mu \varepsilon}} 
\end{equation}
where $\mu = (m_{e}^{*}m_{h}^{*})/(m_{e}^{*}+m_{h}^{*})$ is the reduced mass, $m_{e,h}^*$ denotes the effective masses for electrons ($m_{e}^* = 0.985m_0$) and holes ($m_{h}^* = 0.575m_0$) in Cu$_2$O \cite{walther2020plasma, heckotter2025energy}, $m_0$ the free electron mass, $e$ is the electronic charge, and $\epsilon=7.5\epsilon_{0}$ the permittivity in Cu$_2$O \cite{walther2020plasma, heckotter2025energy}.

The initial velocities of the plasma electrons and holes are sampled from a Maxwell-Boltzmann distribution, where the velocity components follow
\begin{equation}
\eta(v_{e,h}) = \sqrt{\frac{m^*_{e,h}}{2\pi k_B T_{e,h}}} \exp\left(-\frac{m^*_{e,h} v_{e,h}^2}{2k_B T_{e,h}}\right).
\end{equation}
Here, $k_B$ is the Boltzmann constant, similarly $T_{e,h}$ represent the carrier-specific temperatures. 
Unless specified otherwise, our simulations assume thermal equilibrium ($T_e = T_h = T$). The relationship between the plasma temperature and density is characterized by the Coulomb coupling constant
\begin{equation}\Gamma = \frac{e^2 / (4 \pi \varepsilon a_{ws})}{k_B T},\end{equation}
where $a_{ws} = (3/4\pi \rho)^{1/3}$ is the Wigner-Seitz radius, defining the mean interparticle spacing. For neutral plasmas, low recombination rates, $\Gamma < 1$ and large interparticle spacings indicate that quantum correlations are negligible \cite{chen1984introduction}. 
In Rydberg exciton experiments \cite{chakrabarti2025direct, walther2020plasma, kazimierczuk2014giant, heckotter2018rydberg}, typical temperatures range from $1$ to $40$ K with densities spanning $0.001$ to $1$ $\mu$m$^{-3}$. Based on the plasma densities and the corresponding average temperatures reported in the supplementary material of Ref. \cite{heckotter2018rydberg}, we calculated $\Gamma$ to be within the range $[0, 0.2]$. This low coupling constant together with the low plasma densities validate the treatment of the plasma as a system of classical point particles. In our simulations, we initialize the plasma with a Coulomb coupling constant $\Gamma \leq 0.2$, ensuring that the plasma over equilibration is close to an ideal plasma \cite{conde2018introduction}. 
Higher $\Gamma$ values than 0.2 lead to non-negligible recombination rates in our simulations. 



\begin{figure}[h!] 
  \includegraphics[width=1.0\linewidth]{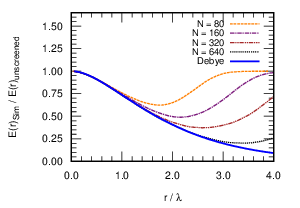} 
\caption{\label{Debye_screening} The ratio between the simulated total electric field to the unscreened point charge electric field as a function of the distance from a central stationary 
point charge. The simulations are performed for increasing number of plasma charges $N$. Also shown is the same ratio for the Debye model. Increasing the number of plasma charges increases the range over which the results are in agreement with the Debye model. The used parameters are $\rho = 0.5\times10^{17}$~m$^{-3}$ and $\Gamma=0.1$.}
\end{figure}

In order to keep the plasma charges confined to a finite volume in our simulations
, we introduce a trap potential in the radial direction $r$. This trap potential is chosen such that it is approximately flat and small in the region of interest around the source charge. This choice means that it has little effect on the plasma charge distribution for a range of $r$ around the origin, as seen in the screening of a point charge in Fig. \ref{Debye_screening}. For a simulation with $N$ plasma particles and plasma density $\rho$, we choose the functional form of the trap potential:
\begin{equation}
    V(r) = k_B T \, \frac{r^\zeta}{\zeta R^\zeta},
\end{equation}
$R$ is the radius of a sphere containing the $N$ particles such that $ \frac{4 \pi}{3} R^3 = N / \rho$. An exponent $\zeta=6$ was enough to accurately simulate Debye screening up to $r\sim R$ for a point charge, cf. Fig. \ref{Debye_screening}.
We also 
performed simulations using higher values of the exponent $\zeta$ and obtained similar results. 



We implemented a fourth-order RKQS \cite{press2007numerical} numerical integrator with an adaptive time step for all simulations. The plasma particles undergo collective interactions driven by mutual electrostatic (Coulomb) forces. Numerical integration of the equations of motion is performed until the plasma reaches thermal equilibrium, at which point the charges are coupled to the system under investigation. 
This `prerun period' is typically taken to be around fifty plasma oscillation periods. To verify that the plasma has reached a steady state, we monitor the local temperature and density by calculating the average kinetic energy and particle count within spheres of radii $R$ and $R/2$. Our results confirm that thermal and spatial equilibrium are established well before the end of the prerun period. 
Those equilibrium temperatures and densities are the ones reported later in the analysis. In our simulations, we use a soft-core $a_s$ in the electric forces to prevent divergences when charges get close. We replace $r_{jj'}$ with $ \sqrt{r_{jj'}^2 + a_s^2}$ in the interaction potential between particles $j$ and $j'$, where $a_s$ was taken to be about $0.02$~a$_{ws}$ \cite{niffenegger2011early}. We tested convergence in the soft-core by performing calculations with $a_s=\{0.01,0.02,0.04\}$~a$_{ws}$ and did not observe a change in the results.

In Fig. \ref{Debye_screening} we show that this model reproduces the known Debye screening for a stationary point charge where the Debye-screened potential $\Phi(r) = \frac{Q}{4 \pi \varepsilon r} e^{-r/\lambda}$ with $\lambda = \sqrt{\frac{\epsilon k_B T}{\rho e^2}} $ being the Debye length. The simulations were performed with equal numbers of holes and electrons, each $=N/2$, at a density $\rho = 0.5\times 10^{17}$~m$^{-3}$ and a Coulomb coupling constant $\Gamma = 0.1$. As can be seen in the figure, by increasing the number of particles in the simulation, the radius up to which screening can be observed increases. Since we are working with a neutral plasma, at radii $(r\gg R)$ 
the electric field averages to that of a point charge, as expected from Gauss's law. We performed simulations for multiple other values of temperature and density, and obtained similar agreement with the Debye screening for a stationary charge at radii $(r<R)$.

\subsection{Modeling the excitons}
\label{subsec_ExcitonModels}
To investigate the effects of a plasma background on Rydberg excitons, we employ two complementary representations of the exciton: a classical Keplerian orbit model ("classical model") and a quantum harmonic oscillator model ("semiclassical model") .

\subsubsection{Classical Model}
\label{Methods_ClassicalModel}
In the classical model, we let the exciton's electron and hole follow fixed classical Keplerian orbits. 
The pair is initialized with energies and relative motion semi-major axis $a_{n}$ matching the hydrogenic states:
\begin{eqnarray}
\label{eq_E_n-a_n}
    E_n &=& -\frac{\mu}{2 \hbar^2} \left(\frac{e^2}{4\pi \epsilon}\right)^2\left(\frac{1}{n^2}\right) \nonumber \\
    &=&-\frac{e^2}{4\pi \epsilon} \frac{1}{2a_n}
\end{eqnarray}
The orbital semi-major axes of each of the electron and hole are determined by their effective masses, with the hole's being $\mu a_n/m_{h}^{*}$ and the electron at $\mu a_n/m_{e}^{*}$. The orbital frequency of these states is given by the classical Rydberg frequency $\omega_{Ryd}$:
\begin{eqnarray}
    \label{eq_classic_freq}
    \omega_{Ryd} &=& \frac{\mu e^{4}}{16\pi^{2}\epsilon^{2}\hbar^{3}} \left(\frac{1}{n^{3}}\right) \nonumber \\
    &=& \frac{2\pi}{t_{Ryd}}.
\end{eqnarray}

For high angular momentum states, the particles are initialized on circular orbits. For lower angular momentum states $l$, we utilize elliptic orbits where the eccentricity $\xi$ is determined by the angular momentum $L = \hbar \sqrt{l(l+1)}$ via \cite{goldstein19801}:
\begin{equation}
\label{eq_ecc}
    \xi = \sqrt{1-\frac{L^{2}}{\mu k_{e}e^{2}}\left(\frac{\mu\omega_{Ryd}^{2}}{k_{e}e^{2}}\right)^{1/3}}
\end{equation}
where $k_{e} = 1/(4\pi\epsilon)$. For typical low-$l$ states ($l=0,1,2$), eccentricities range from 0.9 to 1.0.

In Cu$_2$O, Rydberg exciton states with different orbital angular momentum values possess different energies due to a quantum defect~\cite{kazimierczuk2014giant,schone2016coupled,schone2016deviations}. Consequently, the classical elliptic orbits will exhibit precession. 
For a bound electron-hole pair in a highly elliptical orbit, the precession frequency is proportional to the energy difference between states sharing the same principal quantum number $n$ but differing in $l$ \cite{hezel1992classical}. Thus, when discussing the screening of exciton orbits in Sec. \ref{reslts_screening}, we include a precession frequency for the exciton's orbit.

This model can also be used to obtain a classical estimate of the exciton's state lifetime. By including the back-action of the plasma particles on the exciton charges, the exciton's orbits are perturbed and ultimately broken apart. The lifetime is obtained by tracking the fraction of the trajectories that leaves the energy interval $[E_{n-1}, E_{n+1}]$ as a function of time, after random initialization of the plasma. In this Monte-Carlo scheme, the excitons counts as having decayed, once a trajectory leaves the energy range. Taking the energy window to be this large is a conservative approach that tends to overestimate the lifetimes. The results are obtained by averaging over millions of trajectories, until convergence is reached.

\subsubsection{Semi-Classical model}

\label{methods_HOModel}

To refine our estimate of plasma-induced decay rates, we employ a semi-classical model based on the properties of excitons at high principal quantum number ($n \gg 1$). For these states, the Hamiltonian simplifies according to two physical features. First, the energy spectrum is approximately equidistant, with a separation $\hbar \omega_{Ryd}$ between state $n$ and its neighbors $n \pm 1$. Second, the interaction with the plasma background acts primarily as a dipole coupling that drives transitions between these adjacent levels. We calculate this coupling, $V_{ex-pl}$, as the time-dependent total Coulomb potential exerted by the $N$ classical plasma particles:
\begin{eqnarray}
\label{eq_V_ex_lp}
V_{ex-pl} &=& \sum_{i=1}^{N} \frac{q_{i}}{4\pi\epsilon} \left( \frac{e}{|\vec{r}_{i} - \vec{r}_{h}|} - \frac{e}{|\vec{r}_{i} - \vec{r}_{e}|} \right) \nonumber \\
&\approx& \sum_{i=1}^{N} \frac{q_{i}}{4\pi\epsilon} \frac{e (\vec{r}_{h} - \vec{r}_{e})\cdot\vec{r}_{i}}{|\vec{r}_{i}|^3}
\end{eqnarray}
where $q_i$ and $\vec{r}_i$ denote the charge and position of the $i$-th plasma particle. As shown by the expansion in Eq.~(\ref{eq_V_ex_lp}), the leading term corresponds to a dipole interaction, which couples the state $n$ to the neighboring levels $n \pm 1$. Since plasma induced transitions between the angular momentum states of the same $n$ state are neglected, our analysis gives a conservative estimate of the exciton states lifetimes. 

Given this structure, 
we map the exciton's internal dynamics onto a quantum harmonic oscillator of frequency $\omega_{Ryd}$. In this framework, the classical plasma particles act as a fluctuating environment that induces incoherent transitions between the oscillator levels. To ensure that the magnitude of the oscillator's dipole moment matches that of the Rydberg exciton's, we initialize the system in a state $n_{osc}$ such that the orbit of the harmonic oscillator is the same as that of the hydrogenic state:
\begin{equation}
\label{dipole_mom}
    \sqrt{n_{osc}+1/2}\sqrt{\frac{e^{2}\hbar}{m_e \omega_{Ryd}}} = e\langle r\rangle_{n} = \frac{6\pi\epsilon\hbar^{2}}{\mu e}n^{2},
\end{equation}
where $\langle r\rangle_{n}$ represents the expectation value of the radial coordinate in the state of principal quantum number $n$. This is an approximation for the states of interest, namely Rydberg states with low angular momentum $l \ll n$ states. 
The time evolution of this system is calculated using the Truncated Wigner Approximation (TWA) as detailed in Sec. \ref{methods_wigner_num}. We also demonstrate in App. \ref{app_toy_model}, using a toy model of single charge scattering off an oscillator, that the TWA can reproduce the results obtained by exact 
numerical evolution of the Schr\"{o}dinger equation. We believe that a solution to the Schr\"{o}dinger equation of the electron-hole pair in the presence of a classical plasma would yield reliable predictions in the limit of uncorrelated ideal plasma. This semiclassical model is suited to describe the early-time redistribution of population over the excitonic state due to the plasma's influence.


\subsection{\label{methods_wigner_num}Time evolution of the Wigner function}

We use the Wigner function formalism \cite{wigner1st} that maps the density operator of the excitonic Rydberg state onto a quasi-probability distribution in the phase space via: 
\begin{align}
\label{eq:Wigfun}
W(x, p) = \frac{1}{2 \pi \hbar} \int_{-\infty}^{\infty} dy \, e^{i p y / \hbar} \left\langle x-\frac{1}{2} y|\hat{\rho}(x,y)| x+\frac{1}{2} y\right\rangle .
\end{align}
where $\hat{\rho}(x,y)$ is the density matrix. The Wigner function in this form is real, but not positive definite. 


Conversely, the density matrix elements, mapped onto the basis $\phi_n$, can be obtained from the Wigner function using:
\begin{equation}
\label{den_wig}
    \rho_{nm} = \int_{-\infty}^{\infty} \int_{-\infty}^{\infty} W(x,p) W_{nm}(x,p) dx \, dp ,
\end{equation}
with 
\begin{equation}
    W_{nm}(x,p) = \int_{-\infty}^{\infty} dy \, \phi_m(x+y/2) \phi_n(x-y/2) e^{-i p y/\hbar}.
\end{equation}

The time-evolution of the Wigner function is obtained from the Schr\"{o}dinger equation \cite{wolfgang2011quantum} and takes the form:
\begin{equation}
\frac{\partial W}{\partial t}=\mathcal{T}[W]+\mathcal{U}[W]
\end{equation}
with 
\begin{align}
\label{eqn:wig_tuevo}
    \mathcal{T}[W] &= -\frac{p}{\mu} \frac{\partial}{\partial x} W \notag\\
    \mathcal{U}[W] &= \sum_{l=0}^{\infty} \frac{(-1)^l(\hbar / 2)^{2 l}}{(2 l+1) !} \, \frac{\partial^{2 l+1} U}{\partial x^{2 l+1}} \, \frac{\partial^{2 l+1}}{\partial p^{2 l+1}} W(x, p),
\end{align}
where $\mathcal{T}$ comes from the kinetic part of the Hamiltonian and $\mathcal{U}$ accounts for the portion of the time evolution that is driven by the potential $U$. If the potential energy is a polynomial of order 1 ("constant force") or 2 ("harmonic potential"), the expression for $\mathcal{U}$ in Eq.~(\ref{eqn:wig_tuevo}) takes a simple form:
\begin{equation}
    \mathcal{U}[W] = \frac{\partial U}{\partial x} \, \frac{\partial}{\partial p} W(x, p)
\end{equation}
and the time evolution of the Wigner function becomes: 
\begin{equation}
\frac{\partial W}{\partial t} = -\frac{p}{\mu} \frac{\partial}{\partial x} W + \frac{\partial U}{\partial x} \, \frac{\partial}{\partial p} W(x, p) \label{eq:HO_wig_evo}
\end{equation}
which is identical to the Liouville equation for the classical probability distribution~\cite{Wig-Pedestrians}. For higher order potentials, Eq.~(\ref{eq:HO_wig_evo}) is the truncated  Wigner approximation (TWA) \cite{czischek2020discrete,Nedjalkov2011}. In either case, the phase space probabilities evolve following classical trajectories. In App.~\ref{trace_Wigner}, we outline how to obtain the time evolution of the reduced Wigner function $W_s$ for the exciton state by tracing out the classical plasma degrees of freedom. 



To numerically implement the TWA to our simulations of the exciton (mapped onto a harmonic oscillator) inside the plasma, we represent the continuous Wigner phase-space distribution as a statistical ensemble of discrete classical trajectories. We initialize this ensemble using a Monte Carlo rejection scheme that samples phase-space points $(x_0, p_0)$ proportional to the magnitude of the initial Wigner function, $|W(x,p; t=0)|$. To rigorously account for the non-classical regions where the Wigner function is negative, each trajectory is assigned a weight $w_j = \text{sgn}[W(x_0, p_0)]$ of either $+1$ or $-1$. At each time step, the reduced Wigner function of the exciton is recovered by integrating out the plasma degrees of freedom (see App.~\ref{trace_Wigner}). In this framework, the constituent trajectories propagate according to classical Hamiltonian equations, while the time-dependent Wigner function $W(x,p; t)$ is reconstructed via an ensemble average of the weighted trajectories. This technique is similar to the ones in \cite{shao2015comparison} and \cite{torres2009simulation} in assigning a weight to each Monte Carlo trajectory. As the number of trajectories increases (a few million trajectories in our simulations), this ensemble average converges to the solution of the time-evolved Wigner function.

Since the TWA is only exact for the linear and quadratic potentials, applying it to the case of Coulomb scattering ($\propto \frac{1}{r}$), is not a straightforward assumption. In App.~\ref{app_toy_model}, we numerically solve the Schr\"{o}dinger equation for a single charge wave packet scattering off a harmonically oscillating dipole, showing that, for the typical values of dipoles and separations in our system, the TWA accurately reproduces the time evolution of the state's population obtained from the full quantum mechanical Schr\"{o}dinger equation evolution.



\section{\label{sec_Results} Results}
In this section, we discuss the simulation results pertaining to a single exciton inside the plasma as well as the plasma screening of two interacting excitons using a harmonic oscillator model. In Sec.~\ref{Sec_thermalneo}, the thermalization of an exciton is discussed using the Wigner function of the exciton approximated as a harmonic oscillator with the classical frequency. In Sec.~\ref{reslts_screening}, we use a classical orbit model to discuss the screening between the excitons's electron and hole when embedded inside a neutral plasma. Finally, in Sec.~\ref{results_coupled}, the effects of the plasma screening on the transfer of energy between coupled excitons are discussed. 

\subsection{Lifetimes of exciton states}
\label{Sec_thermalneo}

Extensive experimental research has been conducted on exciton lifetimes \cite{kazimierczuk2014giant,heckotter2025temperature,chakrabarti2025direct}, complemented by theoretical investigations into the specific contributions of semiconductor phonon scattering to these decay dynamics \cite{stolz2018interaction, schweiner2016}. For excitons at small principal quantum numbers, the lifetime $\tau_n$ is observed to scale as $\tau_n \propto n^3$, in analogy to the scaling of radiative decay \cite{elliott1957intensity}.
Beyond $n\simeq 15$, however, the lifetimes often appear to saturate even in high-purity samples.  
Previous studies suggested plasma effects as potential limiting factors for the excitonic lifetimes \cite{semkat2019influence}. 
In this section, we systemically study how such decay channels depend on plasma parameters such as density and temperature as well as the exciton state's principal quantum number. 

An expectation for the trends in the exciton state's lifetime can be developed analytically using an electron scattering off an oscillating dipole. If the plasma particles have a temperature $T$, density $\rho$ and exciton has a scattering cross section $\sigma_n$, then the scattering rate of the plasma particles is $\nu\propto\rho \sigma_n T^{1/2}$. 
Within the first-order Born approximation~\cite{sakurai2020modern}, the scattering cross section of the exciton's long-range dipole potential is proportional to the exciton radius, yielding $\nu\propto\rho \langle r^2\rangle_n T^{-1/2}$ \cite{itikawa1978electron}. 
We further estimate that lifetime of the exciton state is directly proportional to the energy needed for a transition and inversely proportional to the scattering rate by the plasma particles
\begin{equation}
    \tau_n \propto \frac{|\Delta E_n|}{\nu \bar{\eta}(T)},
\end{equation}
where $\Delta E_n \simeq \hbar \omega_{Ryd} \propto \frac{1}{n^3}$, and $\bar{\eta}(T) \propto k_B T$ is the average exchanged energy in a collision.
Thus, the exciton lifetimes are expected show the following scaling behavior
\begin{eqnarray}
\label{eq_lifetime}
    \tau_n &\propto \frac{1/n^3}{\rho n^4 T^{1/2}} \propto \frac{1}{\rho} \frac{1}{n^7} \frac{1}{T^{1/2} }.
\end{eqnarray} 
We numerically test the estimated scaling using the semiclassical model discussed in Sec. \ref{methods_HOModel}. We average over millions of trajectories for each data point, sampling distinct plasma configurations and initial exciton coordinates from the Wigner function corresponding to the exciton at principal quantum number $n$. We checked that this number of trajectories is sufficient for convergence. 
At late times, incoherent repopulation of the initial state (e.g., $n \rightarrow n+1 \rightarrow n$) causes the population dynamics to deviate from exponential decay, resembling a random walk among oscillator states. Thus, the exciton's states lifetimes are obtained by fitting the early-time evolution of the initial state population to an exponential function.

\begin{figure}[h!] 
  \includegraphics[width=1.0\linewidth]{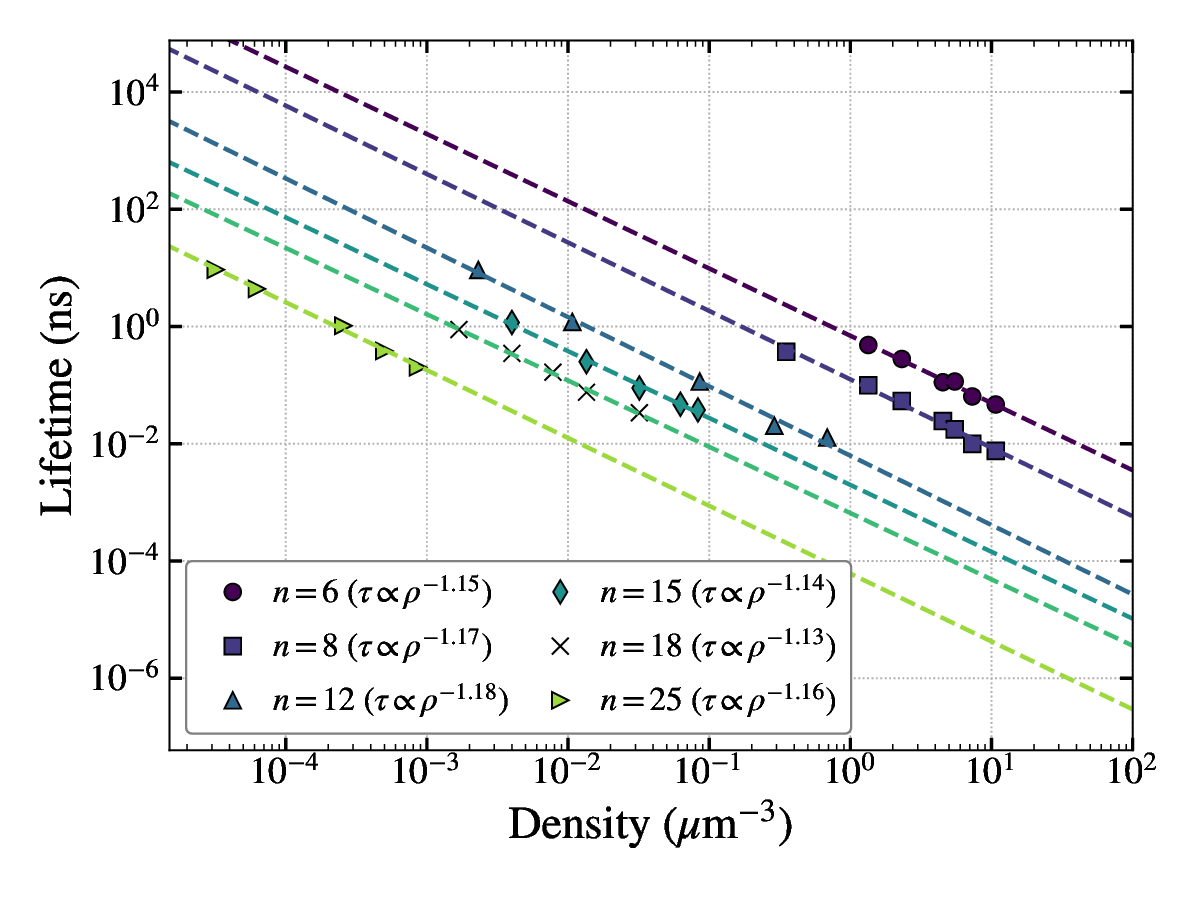} 
\caption{\label{fig_lifetime_vs_den} A log-log plot of the plasma-induced 
lifetime of the exciton state as a function of the density at a temperature of 40~K for different principal quantum numbers $n$. The dashed lines are the fits to exponential decay functions with the exponent given in the legend.}
\end{figure}

In Fig.~\ref{fig_lifetime_vs_den} the lifetimes of multiple exciton states are plotted as a function of the density at a constant temperature of 40~K. 
For all the principal quantum numbers the lifetimes follow an approximate $1/\rho$ dependence on the density as can be seen in the fit curves. This is consistent with the analytic result in Eq.~(\ref{eq_lifetime}).

\begin{figure}[h!] 
  \includegraphics[width=1.0\linewidth]{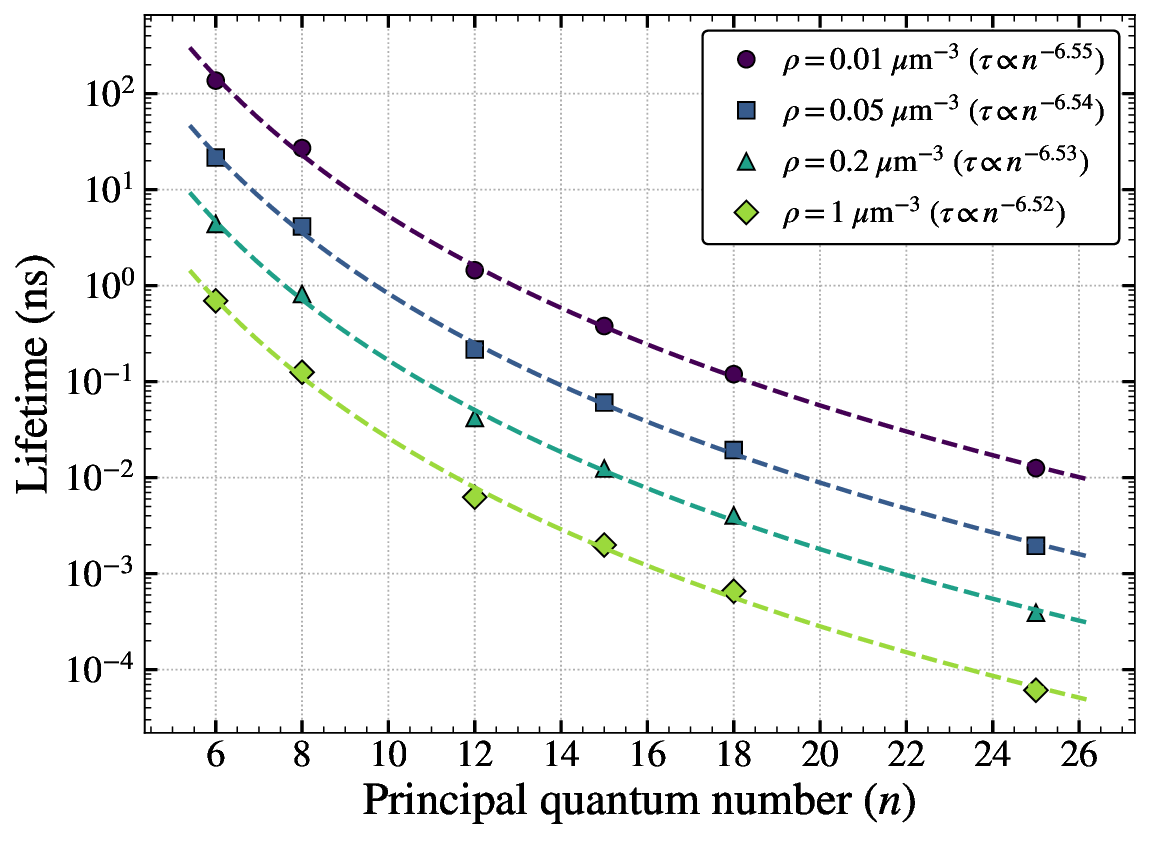} 
\caption{\label{fig_lifetime_vs_n} Exciton lifetimes as a function of principal quantum number, shown for various plasma densities. The data points presented here are extracted directly from vertical cuts of the power-law fit curves shown in Fig. \ref{fig_lifetime_vs_den}.}
\end{figure}

Simulating the cases where the lifetime is significantly longer than the exciton period $t_{Ryd}$ require prohibitively long simulation times. As a result, the feasible density range for simulation varies by multiple orders of magnitude across exciton states. Therefore, to evaluate the dependence of lifetimes on the principal quantum number at fixed densities (Fig.~\ref{fig_lifetime_vs_n}), we take constant-density (vertical) cuts through the \textit{fits} in Fig.~\ref{fig_lifetime_vs_den}. Our results show that the lifetimes indeed decrease with the principal quantum number, approximately matching the $n^{-7}$ expected from Eq.~(\ref{eq_lifetime}). This $\tau_n \propto n^{-7}$ scaling implies a distinctive 'cliff-edge' behavior for observability. As the principal quantum number increases, the scattering rate rises so sharply that states transition from being stable (narrow linewidth) to completely unstable (merging into the continuum) over a very small range of $n$. This effectively limits the observation of a gradual broadening of the peaks for the exciton states \cite{heckotter2018rydberg,kazimierczuk2014giant}.

\begin{figure}[h!] 
  \includegraphics[width=1.0\linewidth]{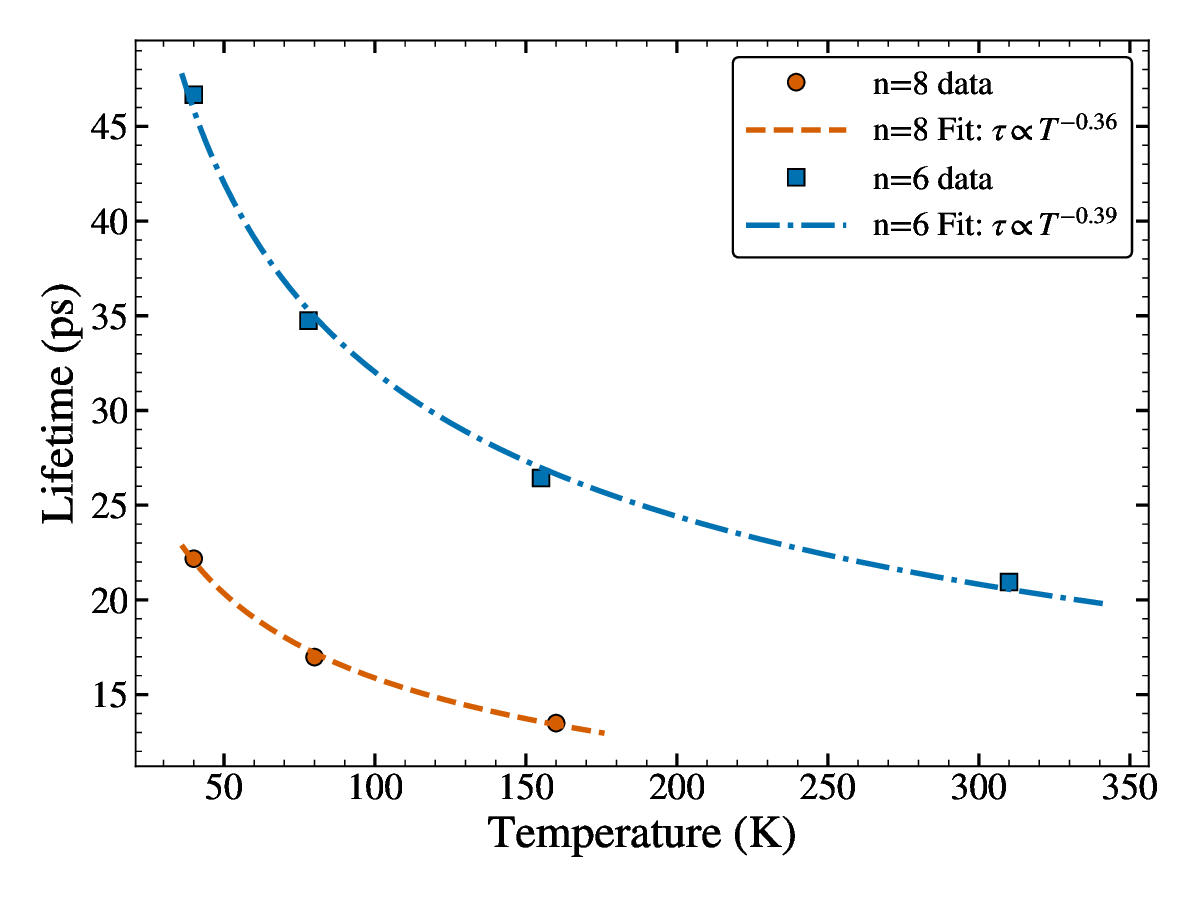} 
\caption{\label{fig_lifetime_vs_T} The lifetime of the exciton states $n=\{6,8\}$ as a function of the temperature at a density of $\{1.07\times10^{19},2.31\times10^{18}\}$~m$^{-3}$ respectively. The marker points are from simulations and the dashed lines are the indicated fits. 
}
\end{figure}

Similarly, by fixing the density and the principal quantum number, the temperature dependence of the lifetime can be determined. As shown in Fig. \ref{fig_lifetime_vs_T}, the lifetime scales approximately as $T^{-1/2}$, in reasonable agreement with the electron-dipole scattering model described in Eq.~(\ref{eq_lifetime}). Simulations were performed for high $n$ states and the lifetime exhibited similar dependence on the plasma temperature. We note that in our simulations, the lifetime becomes less dependent on the temperature as it approaches the Rydberg period ($\tau \sim t_{Ryd}$). 
Simulations were also conducted using non-neutral plasmas, comprised solely of holes or electrons, which yielded lifetimes close 
to those of the neutral plasma. Additionally, in cases where the two plasma species were assigned distinct temperatures, the resulting lifetimes were consistent with the independent summation of the decay rates for each species.


From the dependence of the exciton lifetime on the plasma density (Fig. \ref{fig_lifetime_vs_den}), we define a critical density $\rho_{cr}$ for each exciton state $n$ using the Inglis-Teller limit \cite{10.1093/acprof:oso/9780198530282.003.0007}. 
At such densities $\rho_{cr}$, the plasma-induced decay rate is equal to the frequency separation between adjacent Rydberg states ($(\tau^{cr}_{n})^{-1} = \omega_{Ryd} = 2\pi/t_{Ryd}$)
, causing the exciton levels to merge, rendering their spectral lines indistinguishable. This critical density depends on both the principal quantum number $n$ and the plasma temperature $T$. The analytical dependence of $\rho_{cr}$ can be derived from the previously established scaling relations for lifetime and frequency. By assuming a reference state $n'$ at critical density $\rho'_{cr}$ has a lifetime $\tau^{cr}_{n'} = 1/\omega_{Ryd}$, it follows from Eq.~(\ref{eq_lifetime}) and Eq.~(\ref{eq_classic_freq}) that at a fixed temperature
\begin{eqnarray}
     \frac{\rho'_{cr}}{\rho_{cr}} \frac{n'^7}{n^7} &=&  \frac{\tau^{cr}_n}{\tau^{cr}_{n'}} = \frac{t_{Ryd}}{t'_{Ryd}} \propto\frac{n^3}{n'^3} \nonumber \\
    \rho_{cr} &=& \rho'_{cr} \frac{n'^{10}}{n^{10}} \rightarrow \rho_{cr} \propto \frac{1}{n^{10}}.
\end{eqnarray}
A similar treatment for a fixed state $n$ yields the temperature scaling
\begin{eqnarray}
\label{eq_rho_T}
\rho^{T}_{cr} = \rho^{T'}_{cr}\sqrt{\frac{T'_{cr}}{T_{cr}}}.
\end{eqnarray}

\begin{figure}[h!] 
  \includegraphics[width=1.0\linewidth]{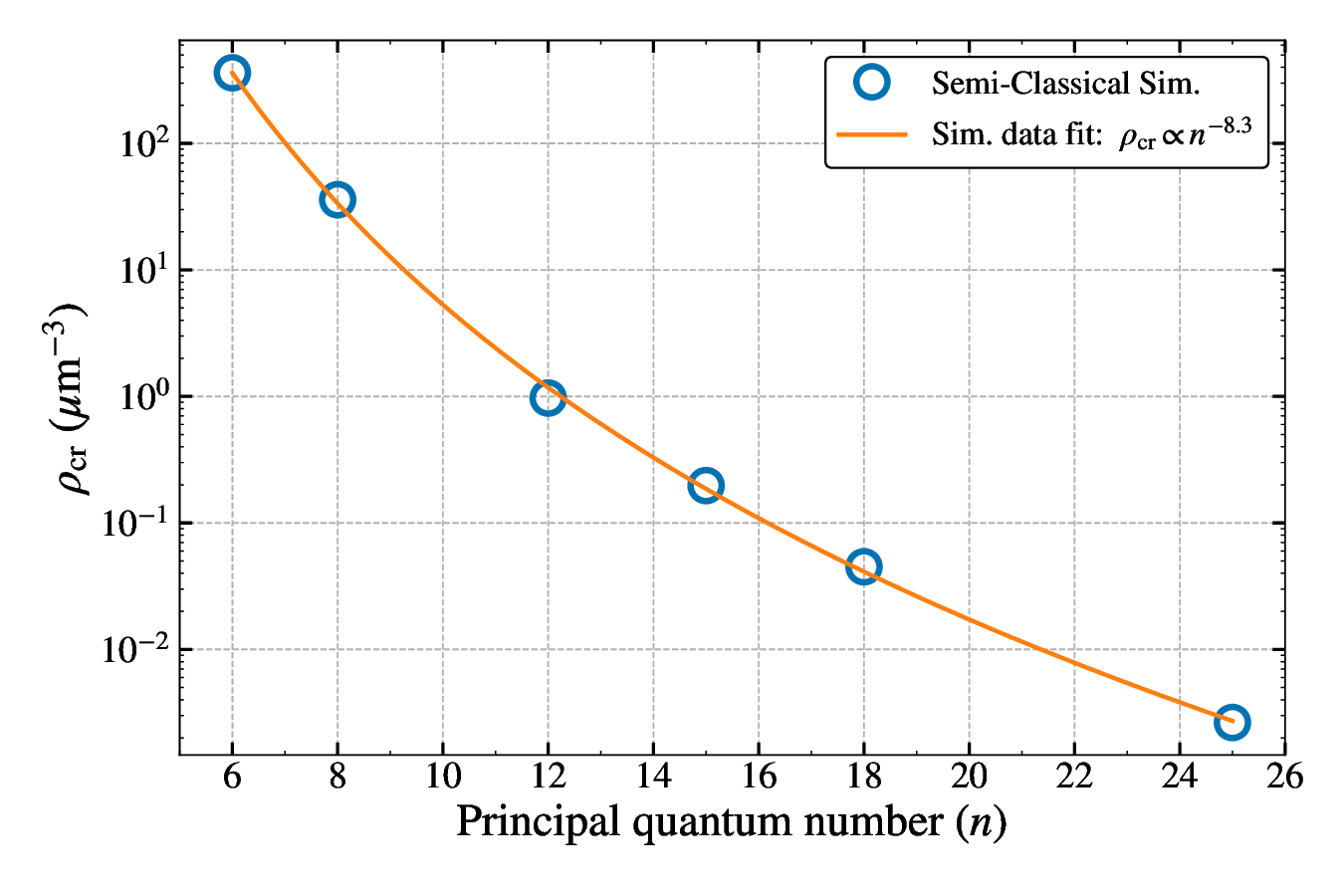} 
\caption{\label{fig_den_vs_n} The critical density beyond which the Rydberg states with principal quantum number $n$ is not distinguishable from its neighboring states at a plasma temperature $T=40$~K. The blue simulation points are obtained using the fit curves in Fig. \ref{fig_lifetime_vs_den} and the orange solid curve indicates the fit. 
}
\end{figure}

Our simulation results for a fixed density and temperature show that the critical density scales as $n^{-8\sim9}$, Fig.~\ref{fig_den_vs_n}, which qualitatively 
aligns with results obtained via the Born approximation. A direct quantitative comparison with Ref.~\cite{heckotter2018rydberg} is complicated by the fact that the plasma parameters reported therein are model-dependent quantities. Specifically, the authors attribute the disappearance of exciton resonances in the absorption spectrum to the Debye-induced shift of the Cu$_2$O band edge. Our analysis, however, suggests that this disappearance is not the result of the band edge crossing the bound states. Rather, our results indicate that the loss of spectral visibility arises from plasma particles scattering off the exciton. This aligns with the authors' subsequent discussion in Ref.~\cite{stolz2022scrutinizing}, which shows the inconsistency of the Debye model. 

\begin{figure}[h!] 
  \includegraphics[width=1.0\linewidth]{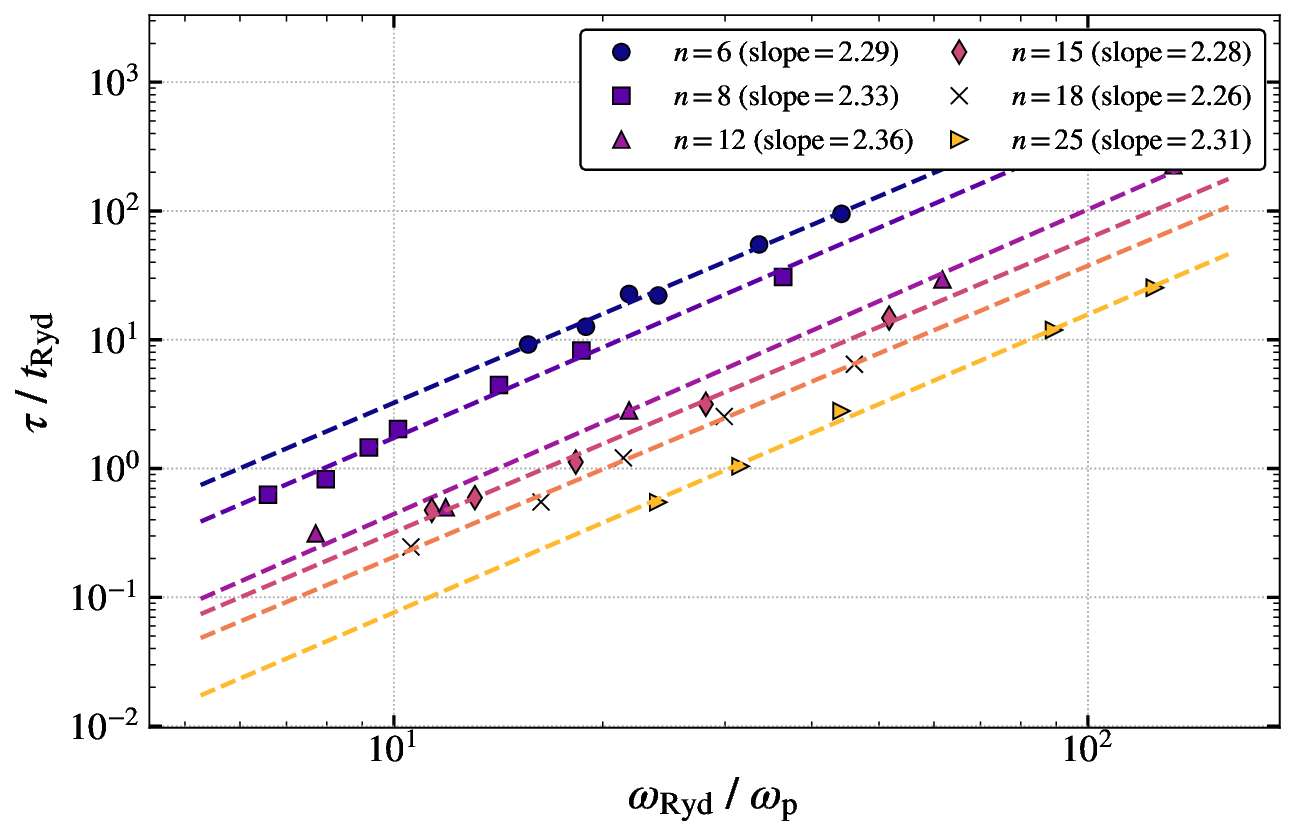} 
\caption{\label{fig_lifetimeRyd_vs_freqRatio} 
Plasma-induced lifetime and against inverse plasma frequency, evaluated at $T = 40$ K. Dashed lines represent power-law fits for each $n$ state and the derived power law exponents. The normalization is with respect to the $n$-dependent Rydberg period $t_\text{Ryd}$ and Rydberg frequency $\omega_\text{Ryd} = 2\pi/_\text{Ryd}$.}
\end{figure}

A key parameter governing the plasma-exciton interaction is the adiabaticity ratio, $\omega_{Ryd}/\omega_p$, which compares the exciton's orbital frequency to the plasma response frequency. This ratio is crucial for understanding the screening effects discussed in Sec.~\ref{reslts_screening}. To explore this relationship, we plot the exciton lifetime normalized by the Rydberg period, $\tau/t_{Ryd}$, against $\omega_{Ryd}/\omega_p$ in Fig.~\ref{fig_lifetimeRyd_vs_freqRatio}. Analytically, combining the lifetime scaling, Eq.~(\ref{eq_lifetime}), with the frequency definitions, Eq.~(\ref{eq_plasma_freq}) and Eq.~(\ref{eq_classic_freq}), yields the relation $\tau/t_{Ryd} \propto (\omega_{Ryd}/\omega_p)^2 n^{-4} T^{-1/2}$. The numerical results in Fig.~\ref{fig_lifetimeRyd_vs_freqRatio} exhibit a slope reasonably consistent with this prediction.

We cross-checked the above results from the semiclassical model of the exciton against the fully classical model (Sec. \ref{Methods_ClassicalModel}). 
Although some quantitative deviations appear, arising from the classical approximation of a quantum system as well as the choice of a wide energy range $[E_{n-1}, E_{n+1}]$, the qualitative dependencies on density, temperature and quantum number are consistent. The classical orbits considered here were elliptic ones with eccentricities, Eq.~\ref{eq_ecc}, corresponding to low angular momentum states, $\xi \gtrsim 0.9$. We note that, higher $l$ orbits tended to have shorter lifetimes at the same plasma density and temperature (lifetimes for $\xi=0.02$ were $\sim20\%$ shorter than those for $\xi=0.95$).

\subsection{Screening of exciton states  inside the plasma}
\label{reslts_screening}

In this section, we examine the extent to which the plasma background screens the Coulomb interaction between the exciton's constituent electron and hole. We employ the classical Keplerian orbit model described in Sec. \ref{Methods_ClassicalModel}. Unlike the previous section, where back-action was included, here the orbital motion of the electron-hole pair is fixed. We do that to isolate the screening effects from the decay dynamics discussed in Sec. \ref{Sec_thermalneo}.

We embed specific classical trajectories of the electron-hole pair within a neutral plasma of density $\rho$ and temperature $T$. We calculate the total electric field along the line connecting the exciton charges and at the location of the hole, generated by both the bound electron and the surrounding plasma charges. We compare this field to that predicted by the standard Debye model:
\begin{equation}
\label{eq_screened_field}
    E(r) = -\partial_r V(r) = \frac{e}{4 \pi \epsilon} \partial_r \left(\frac{e^{-r/\lambda}}{r}\right) 
\end{equation}
The electric field is averaged over thousands of independent simulations until convergence is reached. In each simulation, the plasma is evolved for a pre-run phase of fifty plasma periods ($\tau_{p}=2\pi/\omega_{p}$) to reach a steady state, after which the electric field is averaged over another few plasma periods.

The efficiency of plasma screening is fundamentally governed by the ratio of the exciton's orbital frequency to the plasma frequency, $\omega_{Ryd}/\omega_{p}$, as identified in Sect.~\ref{Sec_thermalneo}. This ratio dictates the adiabaticity of the system: it compares the timescale on which the bound charges move to the timescale required for the plasma background to reorganize and screen them. Based on the analysis in Sec.~\ref{Sec_thermalneo} and prior experimental observations \cite{heckotter2018rydberg}, observable Rydberg exciton states in Cu$_{2}$O typically occupy the high-frequency regime ($\omega_{Ryd} \gg \omega_{p}$), where the orbital dynamics are significantly faster than the plasma response. To comprehensively map the crossover between screened and unscreened behavior, we perform simulations across a wide range of frequency ratios, extending beyond the experimentally typical values to identify the precise limits where the stationary Debye assumption breaks down.

Following the discussion in Sec.~\ref{Methods_ClassicalModel}, the low angular momentum exciton states possess elliptic orbits. According to Eq. \ref{eq_ecc}, these states correspond to eccentricities $\xi$ ranging from 0.9 to 1.0. We present results here for a representative eccentricity of $\xi=0.95$; simulations using other eccentricities within the above range yielded similar results. Furthermore, as discussed in Sec.~\ref{Methods_ClassicalModel} these elliptical orbits exhibit precession due to the the quantum defect. We use an approximate precession frequency of one-fifth of the classical orbital frequency $0.2\times\omega_{Ryd}$, corresponding to the defect between the $l=0$ and $l=1$ of the Rydberg exciton states \cite{kazimierczuk2014giant,schone2016coupled,schone2016deviations}. In App.~\ref{Screening_circle} we discuss the screening of high angular momentum states (approximately circular orbits).

Along the elliptic orbit, the separation between the two charges varies from a minimum of $a_n(1-\xi)$ to a maximum of $a_n(1+\xi)$
, where $a_n$ is defined in Eq.~(\ref{eq_E_n-a_n}). To quantify the electric field screening, we discretized this range into small intervals and compared the average total electric field within each interval to the Debye-screened field at the interval's midpoint.

\begin{figure}[h!] 
  \includegraphics[width=1.0\linewidth]{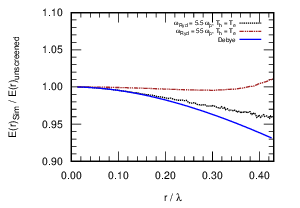} 
\caption{\label{fig_pr_ellipse_diffmass} 
Ratio of the simulated electric field, at the location of the hole, for an electron-hole pair in an elliptic orbit to the unscreened point charge field as a function of the radial separation $r$. The dotted black curve corresponds to the screening of the $n=20$ state at a plasma density of $\rho = 6.25\times10^{16}$~m$^{-3}$ and a plasma temperature of $T=7$~K. The dash-dotted red curve corresponds to the same plasma density and temperature but with an exciton frequency $10\times$ higher.
}
\end{figure}

Figure~\ref{fig_pr_ellipse_diffmass} illustrates the transition from partial to negligible screening by varying the frequency ratio $\omega_{Ryd}/\omega_p$ for an exciton state. In the regime where the plasma density is close to the state's critical density ($\rho \approx \rho_{cr}$, corresponding to $\omega_{Ryd} = 5.5\omega_p$), the plasma partially tracks the exciton's precessing dipole, and the static Debye model only slightly overestimates the effective screening. We confirmed that this behavior is insensitive to the initial thermal distribution of the plasma; simulations initialized with disparate electron and hole temperatures ($T_h=12$~K and $T_e=2$~K) yielded identical screening profiles. We note that the electric field averaging time window was about $10~\tau_p \sim 10$~ns (comparable to the plasma lifetime window reported in the supplementary material of \cite{heckotter2018rydberg}). By the end of that window, rapid inter-species scattering had brought the electron and hole temperatures to near thermal equilibrium ($T_h=8$~K and $T_e=6$~K). At higher Rydberg-to-plasma frequency ratios, the electric field remains essentially unscreened, as the plasma cannot respond to the rapidly rotating charges.

Other work has been presented in the literature to solve for the plasma screening effects on the exciton's internal interaction using a quantum many-body approach \cite{haug1984electron,stolz2022scrutinizing}. Such treatments were performed with approximations such as; low density or quasi-static approximations. These approximations limit the regime of applicability of the results obtained near the $\omega_p\sim\omega_{Ryd}$ regime and make them a first order correction to the Debye static approximation in the regime $\omega_p <\omega_{Ryd}$. Our results in this section do not aim to provide a quantitative result of the interaction. We only aim at giving a simpler picture to identify which regimes should one expect the Debye model to be a close enough approximation and which ones it is not. We note that from our results one would expect that higher order corrections to the Debye model would reduce the screening effects. This agrees with the results in \cite{semkat2019influence} for the low $n$ states. However, the energy shifts calculated in \cite{semkat2019influence} for the high $n$ states (i.e the quantum corrections to the Debye model leading to larger energy shifts) are contrary to our simpler picture expectations and we think further analysis of the applicability of a first order perturbative approach for the $\omega_p\sim\omega_{Ryd}$ regime could be needed. We also note that, generally the critical densities expectations from our scattering model in Sec.~\ref{Sec_thermalneo} are lower than the ones expected from the Debye band gap shift model \cite{heckotter2018rydberg, semkat2019influence}.


\subsection{Regimes of screening and observability}
\label{Sec_thermScreen}

In this section, we combine the screening results in Sec.~\ref{reslts_screening} and the Lifetimes analysis in Sec.~\ref{Sec_thermalneo} to quantify the amount of observable electric field screening before a Rydberg state becomes indistinguishable in the absorption spectrum. As illustrated in Fig.~\ref{fig_lifetimeRyd_vs_freqRatio}, highly excited Rydberg states ($n>10$) become spectroscopically indistinguishable from their neighbors when the Rydberg frequency drops below approximately four times the plasma frequency ($\omega_{Ryd} \lesssim 4\omega_p$). 
This observation can be translated into a statement about the critical ratio between exciton size $\langle r\rangle_{n}$ and Debye length $\lambda$ 
\begin{equation}
\label{r_lambda}
    \frac{\langle r\rangle_{n}}{\lambda} = 2^{1/3} 3^{1/6} \sqrt{\Gamma} \left(\frac{\omega_p}{\omega_{Ryd}}\right)^{2/3}.
\end{equation}
For example, if we assume $\Gamma=0.2$ and a frequency ratio $\frac{\omega_p}{\omega_{Ryd}}=0.25$, we find $\frac{\langle r\rangle_{n}}{\lambda} = 0.18$. It follows that the electric field Debye screening, Eq.~(\ref{eq_screened_field}), at $\langle r\rangle_{n}$ becomes less than 5\%. And our simulation results indicate that the Debye model tends to overestimate the screening. 
This suggests that, except for an overall shift of all the states \cite{ecker1956zustandssumme}, the energy difference between the exciton states and their wavefunctions' radial extent experience limited change by the plasma until the highest observable Rydberg state. 

\begin{figure}[h!] 
  \includegraphics[width=1.0\linewidth]{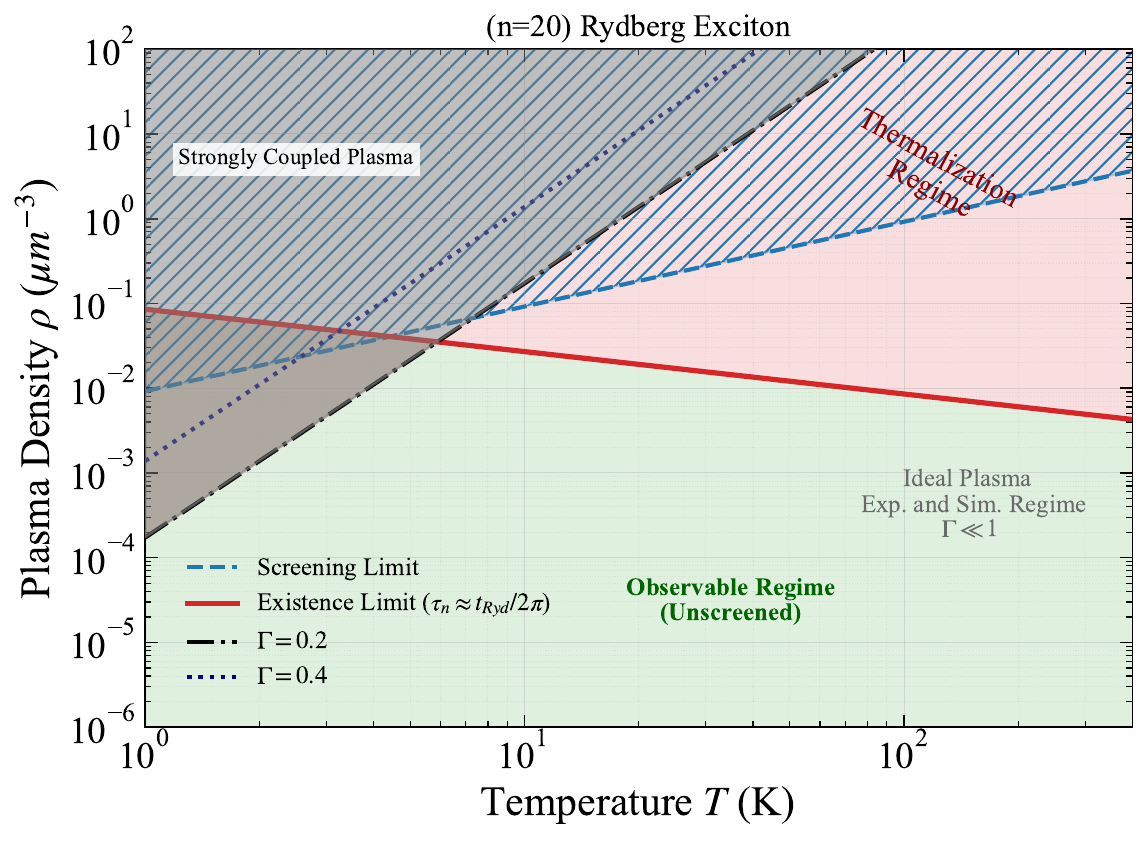} 
\caption{\label{fig_regime_diagram} A density-temperature ($\rho-T$) diagram for the Rydberg exciton at $n=20$. The solid red line delimits the observable regime in which the state is observable, and represents the critical density for the exciton state as a function of the temperature, Eq.~(\ref{eq_rho_T}). The dashed blue line marks the screening region with $\frac{\langle r\rangle_{n}}{\lambda}>0.5$ , Eq.~(\ref{r_lambda}), leading to possible observable screening effects. Crucially, the screening region (blue hatching) lies almost exclusively above the existence limit. The grey shaded area indicates a coupled plasma ($\Gamma > 0.2$) which is outside both the experiments and the simulations regimes. 
} 
\end{figure}

To quantify the extent of plasma screening effects on exciton states that remain observable against thermalization, we map the previous two sections results onto the density-temperature ($\rho-T$) plane for a typical Rydberg state ($n=20$, Fig.~\ref{fig_regime_diagram}). We define two critical boundaries to assess the plasma effects. The first is the `Observable Limit' (solid red line), which marks the critical density (Eq.~\ref{eq_rho_T}) above which the exciton state becomes spectroscopically indistinguishable from its neighbors. The second is the `Screening Limit' (dashed blue line), defined as the threshold density required for Debye screening (which is $\sim\times2$ the simulation's screening near the observability threshold) 
to induce a measurable modification of the exciton's structure. Quantitatively, this threshold corresponds to a relative modification of the wave function exceeding $5\%$ 
or exceeding $10\%$ shift in the adjacent energy level splitting. 
Note that we use the splitting between neighboring levels as our metric rather than the absolute energy shift from the ground state \cite{ecker1956zustandssumme,stolz2022scrutinizing}; the latter can exceed 40\% \footnote{Values are obtained by numerically solving the Schr{\"o}dinger equation for a Debye screened Coulomb potential \cite{walther2020plasma,ecker1956zustandssumme}.}. As shown in Fig.~\ref{fig_regime_diagram}, the Screening Limit lies above the Observable line in the weakly coupled regime. Thus, stable excitons only exhibit limited screening before being indistinguishable in the absorption spectrum peaks. This is in agreement with the results in \cite{stolz2022scrutinizing,semkat2019influence} where the observable energy shifts are approximately $\lesssim3\%$ for the observable states. However, we note that extending the analysis to the moderately coupled regime (lower T, higher $\rho$) 
enhances the screening efficiency, Eq.~(\ref{r_lambda}), potentially opening a narrow parameter window where screening effects become more pronounced before the state is unobservable.

\subsection{ Coupled oscillators inside Plasma}
\label{results_coupled}
In this section, we study the effects of a plasma environment on the interaction between two dipoles. We use the semi-classical model, Sec \ref{subsec_ExcitonModels} to represent both excitons. We investigate whether the plasma background screens the electric field between interacting excitons. In order to achieve that, we study the rate of energy transfer from one dipole to the other. Without the plasma, we assume that the dipoles interact via the interaction Hamiltonian
\begin{equation}
    \mathcal{H}_{int} = - k \frac{q_1 q_2}{D^3} z_1 z_2,
\end{equation}
where $\{q_1, z_1\}$ are the charge and position of the first oscillator, respectively, and similarly for the second oscillator with subscript 2 and $D$ is the separation between the two dipoles. If the plasma screens the field from one oscillator at the location of the other one, the time needed for energy transfer will increase.

\begin{figure}[h!] 
  \includegraphics[width=1.0\linewidth]{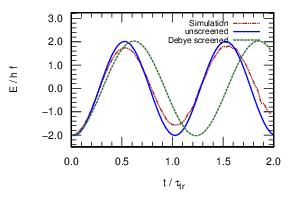} 
\caption{\label{fig_2osc_eng_trans} The energy difference between two coupled excitons embedded inside a plasma as a function of time. One oscillator was initialized in the Rydberg states $n\sim8$ and the other in the ground state. The excitons were separated by $\sim0.8\lambda$ and the plasma density was $0.5\times10^{17}$~m$^{-3}$, giving 
$\omega_{Ryd} = 2\pi f = 50 \omega_p$. The scaled energy difference from the simulation agrees with that of two unscreened dipoles rather than the screened dipoles case. 
} 
\end{figure}

In Fig.~\ref{fig_2osc_eng_trans}, we analyze the energy transfer dynamics between two spatially separated resonant dipoles. One oscillator is initialized in an excited Rydberg state, using the semi-classical model, Sec.~\ref{methods_HOModel}, while the other starts in the ground state. To isolate the coherent dipole-dipole interaction from the incoherent plasma-induced thermalization, Sec.~\ref{Sec_thermalneo}, we normalize the time-dependent energy difference by compensating for the exponential thermalization decay. We then benchmark the simulated energy transfer rate against two theoretical limits: the static Debye-screened interaction and the completely unscreened vacuum interaction. The theoretical transfer rate for the Debye-screened case is calculated using the stationary screened potential~\cite{ugalde1996bound}:
\begin{equation}
\label{eq_scr_dipole_pot}
\Phi(x,z) = \frac{\vec{p}\cdot\vec{r}}{4 \pi \varepsilon r^2}\exp \left( -\frac{r}{\lambda} \right) \left( \frac{1}{r} + \frac{ 1}{\lambda} \right)
\end{equation}
where $r=\sqrt{x^2+y^2+z^2+a_s^2}$ and $a_s$ denotes the softening parameter (see Sec.~\ref{methods_plasma_num}). Our simulations reveal that the energy transfer rates are consistent with the absence of screening. We performed simulations for other Rydberg exciton initialization as well as other separations (order of $\sim \lambda$) between the oscillators with different plasma parameters and observed similar results. This result follows directly from the analysis in Sec.~\ref{reslts_screening}: for fast-oscillating Rydberg states ($\omega_{Ryd} \gg \omega_p$), the plasma responds only to the time-averaged charge distribution. Consequently, the plasma cannot screen the instantaneous oscillating dipole field that drives the resonant energy transfer.

\begin{figure}[h!] 
  \includegraphics[width=1.0\linewidth]{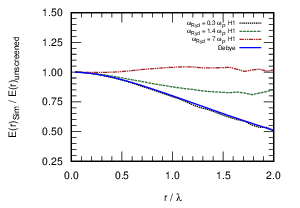} 
\caption{\label{fig_osc_dipole_screening} Ratio of the total electric field to the unscreened Coulomb field as a function of the radial separation $r$ from a central oscillating dipole along the z-direction. Results are shown for the dipole frequencies $\omega = \{0.2, 1.0, 5\}\sqrt{2}\omega_p$ and time-averaged over a half of the oscillation period. The solid curve represents the theoretical Debye screened field in the z-axis direction (from Eq.~\ref{eq_scr_dipole_pot}). The simulation electric field curves averaged over the other half period are similar to the ones shown here.} 
\end{figure}

The semi-classical model employed previously is limited to the cases when the exciton lifetime exceeds the energy transfer period. This constraint effectively restricts this analysis to the high-frequency regime ($\omega_{Ryd} \gg \omega_p$), where plasma-induced decay is slow. To investigate screening across a broader frequency range, including regimes where decay would otherwise occur, we utilize a classical oscillating dipole model. We place a central point dipole oscillating along the $z$-axis at the origin and compute the electric field along the radial direction, $r$. In this approach, the dipole oscillation is fixed rather than dynamical. This effectively suppresses the backaction of the plasma on the exciton, allowing us to isolate and quantify the screening efficiency even in regimes where $\omega_{Ryd}\sim\omega_p$. 
To quantify the plasma's ability to track and screen the changing field, we time-average the electric field over half-periods of the dipole oscillation. The results, Fig.~\ref{fig_osc_dipole_screening}, corroborate our findings from Sec. \ref{reslts_screening}: the plasma can screen the field when its oscillation frequency is lower than the plasma frequency ($\omega_{Ryd} \lesssim \omega_p$). Conversely, no screening is observed when ($\omega_{Ryd} \gg \omega_p$), consistent with the semi-classical simulation analysis.


\section{Conclusion}
\label{sec:Conclusion}

In this work, we have studied the influence of a neutral electron--hole plasma on Rydberg excitons in cuprous oxide, with particular emphasis on plasma-induced lifetime broadening and on the applicability of Debye screening in experimentally accessible regimes. By combining a truncated Wigner description of exciton dynamics with classical simulations of plasma, we have analyzed how plasma scattering, screening, and inter-excitonic coupling manifest under realistic conditions.


A central result of this study is the characterization of plasma-induced thermalization as an important mechanism limiting the lifetimes and observability of highly excited exciton states. Our simulations show that scattering from the plasma leads to mixing of Rydberg levels and a corresponding decay of exciton states, with a strong dependence on plasma density, temperature, and principal quantum number, $\tau_n \propto \rho^{-1}n^{-7}T^{-1/2}$. For small $n$ states, the  lifetimes obtained here are much longer than those caused by phonon-induced channels. Thus, the lifetimes dependence on $n$ is dominated by their $n^3$ scaling for small $n$. However, as $n$ increases, the plasma-induced decay lifetimes become comparable to those due to phonon coupling \cite{stolz2018interaction} and thus contribute to exciton decay. The plasma effects could contribute to the mismatches from the $n^3$ scaling observed in experimental studies \cite{chakrabarti2025direct, kazimierczuk2014giant} on the exciton states' lifetimes at higher $n$.

By explicitly calculating time-averaged electric fields along trajectories of bound excitons, we assess the limitations of applying Debye screening to interactions involving nonstationary excitonic charges. Our results show that, at high $\omega_{Ryd}/\omega$ ratios, the plasma responds primarily to the time-averaged charge distribution of the bound electron-hole pair. For such exciton states our simulations showed that the Debye screening can significantly overestimate the excitonic interaction screening. 


Synthesizing our results on plasma-induced decay and dynamical screening reveals an incompatibility between the conditions required for significant Debye screening and those necessary for exciton resolvability. Our analysis reveals that the plasma density needed to significantly modify the exciton's internal field consistently exceeds the critical density for spectroscopic resolvability. Thus, within the parameter space where exciton states are observable, the extent of screening is intrinsically limited: the plasma density remains too low to trigger perturbations exceeding (5~$\%$ in wavefunction extent or 10~$\%$ in level splitting) before the state is rendered indistinguishable. 
The central approximation in our approach lies in treating the plasma as a collection of classical particles. As noted in the Sect.~\ref{sec:level2}, a full solution of the problem should be obtained from the Bethe-Salpeter equation of the screened electron-hole pair. It is interesting to note that previous work on this topic \cite{semkat2021quantum} found no significant broadening effects on the Rydberg states. An opportunity for future work lies in the exploration of the origins of this discrepancy.

We further examined the influence of the plasma on exciton-exciton interactions by studying energy transfer between coupled excitons. Consistent with the single-exciton analysis, we find no evidence for substantial plasma-induced suppression of inter-excitonic coupling in the regime where individual exciton states are spectroscopically resolvable. This supports the persistence of strong dipole-dipole interactions and blockade effects \cite{kazimierczuk2014giant,heckotter2021asymmetric} under experimentally relevant plasma conditions. 

Finally, we note that our results are strictly conclusive within the weakly coupled plasma regime ($\Gamma \ll 1$). In the strongly coupled regime, which is outside the current experimental regime \cite{kazimierczuk2014giant,heckotter2018rydberg,stolz2022scrutinizing}, quantum many-body treatments are required to accurately capture potential screening phenomena \cite{stolz2022scrutinizing}.



\begin{acknowledgments}
We thank Hadiseh Alaeian for valuable discussions and feedback. AKA and FR were supported by the National Science Foundation under Award No.~2410890-PHY. VW was supported by the National Science Foundation (NSF) under Award $\#$PHYS-2409630. AKA thanks the Rosen Center for Advanced Computing (RCAC) at Purdue University for providing the supercomputing clusters essential for the majority of the plasma simulations performed in this work.
\end{acknowledgments}

\appendix

\section{Tracing out bath in Wigner function evolution}
\label{trace_Wigner}
The initial full Wigner function for the oscillator (coordinates $\{x,p\}$) and the N classical plasma particles (coordinates $\{Q,P\}$) is:
\begin{equation}
    W_{full} (t=0) = W_s(x_0,p_0;t=0) W_p(Q,P;t=0)
\end{equation}
where $W_s$ is the reduced Wigner function for the oscillator and $W_p$ is the plasma particles phase-space distribution obtained from an initial flat distribution inside a sphere for $Q = \{\vec{r}_1, \vec{r}_2,\vec{r}_3...\vec{r}_N\}$ and from the Maxwell-Boltzmann distribution for $P$. Since the plasma is treated as a classical one, therefore its Wigner function is everywhere positive. Thus, the sign of $W_{full}(0)$ is the same as that of $W_s(0)$ and each trajectory is assigned a (+1) or a (-1) depending on the sign of dipole's Wigner function at the initial coordinates $(x_0,p_0)$. The dipole's reduced Wigner function is later obtained from the full Wigner function at each output time $t_p$ by integrating out the plasma particles' degrees of freedom:
\begin{equation}
    W_s(x,p;t=t_p) = \int dQ dP \ \ W_{full}(t=t_p)
\end{equation}
That is achieved by averaging the results from millions of trajectories; each with different initialization of the plasma and the dipole coordinates.  


\section{Agreement between the TWA and SE evolution for scattering}
\label{app_toy_model}

In this section, we discuss a toy model designed to benchmark the Truncated Wigner Approximation (TWA) against a numerical solution of the quantum mechanical Schr\"{o}dinger equation (SE). The system will be a harmonic dipole at the origin and a charge at a substantial distance from the origin. Using a leap-frog algorithm to evolve the SE wave functions, we demonstrate that the TWA accurately reproduces the full quantum dynamics. This agreement justifies our application of the TWA for the more complex simulation of multiple plasma particles scattering off the exciton even though the interaction of a plasma particle with the harmonic oscillator is not a linear or quadratic function of the coordinates.

\begin{figure}[h!] 
  \includegraphics[width=1\columnwidth]{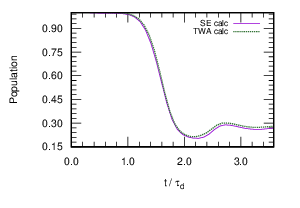} 
\caption{\label{fig_Toy} The population in the initial oscillator's state as a function of time. The time is scaled by the oscillation period. The SE and TWA results agree with each other, thus validating the use of the TWA in the simulations in the text.} 
\end{figure}

The Hamiltonian for the system is given by:
\begin{eqnarray}
        \mathcal{H} &=&\frac{p_1^2}{2m_1} + \frac{p_2^2}{2m_2} + \frac{1}{2} m_1 \omega^2x_1^2  \\&+& k q_1 q_2 \left(-\frac{1}{R + x_1 + x_2} + \frac{1}{R - x_1 + x_2} - \frac{2x_1}{R^2}\right) \nonumber\label{toy_hamiltonian} 
\end{eqnarray}
where the position (x), mass (m), charge (q) and momentum (p) parameters are subscribed by 1 for the exciton and 2 for the point charge. The masses of the particles were taken to be equal to each other and equal to the electron mass. The oscillator's frequency was taken to be equal to $\omega_{Ryd}$ for n$\sim$6. 
The separation $R$ is a function of time and describes the center of mass motion of the point charge as it scatters from the dipole. The extra term $\frac{2x}{R}$ is added to remove the trivial non-coupling first-order interaction term in the potential. The initial separation of the particles, $R(0)$, is set large enough such that the electric interaction is negligible compared to the restoring force of the oscillating exciton such that R(t) is given by: 
\begin{equation}
R(t) = \sqrt{R_c^2+ \left(1-\frac{2t}{t_f}\right)^2\left(R_f^2 - R_c^2\right)}
\end{equation}
with the time starting at $t=0$ and ending at $t_f\sim3.5\times\frac{2\pi}{\omega}$. The initial separation $R_f = 100 d_h$, $d_h=\sqrt{\frac{\hbar}{m_1 \omega}}$, was chosen to be comparable with $a_{ws}$ for a plasma density $\sim1\mu m^{-3}$. The closest approach radius was taken to be $R_c=20d_h\sim0.2a_{ws}$, where we have used a softening parameter for the electric interaction $a_s = 0.02 a_{ws}$. To perform the Schr\"{o}dinger equation simulation, the point charge wavefunction, $\psi(x_2)$, is initialized as a Gaussian wave-packet and the exciton state, $\psi(x_1)$, is initialized in a harmonic oscillator eigenstate, with quantum number n$_{osc}$ obtained from Eq.~(\ref{dipole_mom}), such that at $t=0$ the full system's state is:
\begin{equation}
    \Psi(x_1,x_2;0) = \psi(x_1) \psi(x_2)
\end{equation}
Similarly, the TWA calculation was initialized with the Wigner function of each respective state. We compare the time evolution of the dipole state as a function of time as the point charge scatters off of it from the two simulations. In Fig. \ref{fig_Toy} we see that the two approaches give matching results. For the TWA simulations we averaged a few million trajectories, similar order of magnitude of the number of trajectories of each of the results in the text, and saw a good agreement between the TWA result and the SE result. 

\section{High angular momentum states screening}
\label{Screening_circle}

Here we study the screening of the high angular momentum exciton states. The classical representation of these states are approximately circular orbits. First we consider the simplified case where the electron and hole have equal effective masses. This is relevant for excitons in various other semiconductors \cite{schuster2016nongeneric,wickramaratne2014electronic} and serves as a control study to build intuition for the general plasma screening of the exciton states. The particles are initialized on diametrically opposite sides of a circular orbit. 
The electric field at different separations between the bound electron and hole pair is obtained by considering circular orbits with varying radii. This allows us to determine the electric field as a function of the charge separation distance, $r$.

\begin{figure}[h!] 
  \includegraphics[width=1.0\linewidth]{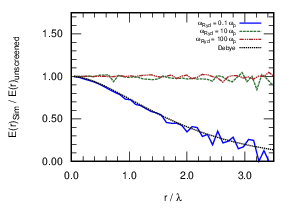} 
\caption{\label{fig_circ_equalmass} The ratio between the simulated electric field, at the position of the hole, to the unscreened point charge electric field as a function of the separation between equal-mass charges rotating on a circle evaluated at frequencies $\omega_{Ryd} = \{0.1, 10, 100\}\omega_p$. The dotted curve represents the statically screened field of the Debye model.
} 
\end{figure}

As shown in Fig. \ref{fig_circ_equalmass}, the screening behavior is significantly different for the different ratios of the orbital frequency to the plasma frequency. 
When the rotation is slow ($\omega_{Ryd} \ll \omega_{p}$), the plasma particles can adiabatically track the moving charges, resulting in standard Debye screening. However, in the high-frequency regime ($\omega_{Ryd} \gg \omega_{p}$), the plasma cannot respond to the instantaneous positions of the electron and hole. Instead, it responds to the time-averaged charge distribution. For equal masses rotating on the same circle, this average distribution is a neutral ring; consequently, the plasma effectively sees zero net charge, and no screening is observed. This has serious implications because, as mentioned in the text, the observable Rydberg states fall into the regime $\omega_{Ryd}\gg\omega_p$. 


\begin{figure}[h!] 
  \includegraphics[width=1.0\linewidth]{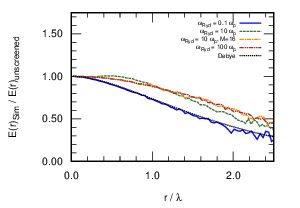} 
\caption{\label{fig_circ_diffmass} Electric field screening for asymmetric orbital radii, see caption of Fig.~\ref{fig_circ_equalmass}. The hole and electron orbit at radii equal to $\frac{\mu r}{m^*_h}$ and $\frac{\mu r}{m^*_e}$, respectively. The dot-dashed curve shows the field resulting from a discretized ring model (16 fragments with charge $\pm e/16$ rotating at $10\omega_p$), corresponding to the time-averaged charge distribution.
} 
\end{figure}

Next, we apply this model to the case of the effective masses of the electrons and holes in Cu$_{2}$O.  
Due to the mass imbalance, the electron and hole orbit at different radii ($\frac{\mu r}{m_{e}^{*}}$ and $\frac{\mu r}{m_{h}^{*}}$), creating two distinct concentric shells of time-averaged charge. The resulting screening profiles, shown in Fig.~\ref{fig_circ_diffmass}, show that at frequencies lower than $\omega_p$, regular Debye screening is observed, whereas the screening is strongly suppressed at higher Rydberg frequencies. 


To rigorously verify that the plasma is responding to the time-averaged charge density rather than the particle's instantaneous position, we performed simulations where the exciton charge was physically distributed along the orbit. We divided the electron (and hole) charge into $M$ equal fragments of magnitude $e/M$, initialized with uniform time separations along the trajectory (corresponding to $\Delta t = \frac{1}{M}\frac{2\pi}{\omega_{Ryd}}$). For fragment counts ranging from $M=2$ to 128, the resulting screening profile was identical to that of the single rapidly orbiting charge, Fig. \ref{fig_circ_diffmass}. This confirms that for fast Rydberg orbits, the plasma effectively responds to and screens the time-averaged charge distribution of the trajectory.


\newpage 
\bibliographystyle{unsrt} 
\bibliography{bibliography} 

\end{document}